\documentclass[preprintnumbers,10pt,nofootinbib]{revtex4}

\usepackage{amsmath,latexsym,amssymb,amsfonts}
\usepackage[pdftex]{color,graphicx}
\usepackage{bm}
\usepackage[normalem]{ulem}

\definecolor{patriarch}{rgb}{0.5, 0.0, 0.5}
\definecolor{darkraspberry}{rgb}{0.53, 0.15, 0.34}
\definecolor{brinkpink}{rgb}{0.98, 0.38, 0.5}
\definecolor{skobeloff}{rgb}{0.0, 0.48, 0.45}
\definecolor{mypink}{RGB}{226,68,130}
\definecolor{darkpastelgreen}{rgb}{0.01, 0.75, 0.24}
\definecolor{pigmentgreen}{rgb}{0.0, 0.65, 0.31}

\begin{document}


\title{\textbf{Unitary evolution of an evaporating black hole in canonical quantum gravity}}

\author{
Sijia Wang$^{a}$\footnote{{\tt s676wang@uwaterloo.ca}}
}
\author{
Robert B.\ Mann$^{a,b}$\footnote{{\tt rbmann@uwaterloo.ca}}
}
\author{
Dong-han Yeom$^{c,d,e}$\footnote{{\tt innocent.yeom@gmail.com}}
}

\affiliation{
$^{a}$Department of Physics and Astronomy, University of Waterloo, Waterloo, Ontario N2L 3G1, Canada\\
$^{b}$Perimeter Institute, Waterloo, Ontario N2L 2Y5, Canada\\
$^{c}$Department of Physics Education, Pusan National University, Busan 46241, Republic of Korea\\
$^{d}$Research Center for Dielectric and Advanced Matter Physics, Pusan National University, Busan 46241, Republic of Korea\\
$^{e}$Leung Center for Cosmology and Particle Astrophysics, National Taiwan University, Taipei 10617, Taiwan
}

\begin{abstract}
We propose a framework for understanding the information loss paradox in the context of canonical quantum gravity. We first revisit several approaches to black hole evaporation that do not involve an event horizon. In these models, we assume that each time slice corresponds to a coherent state in the gravitational phase space. Due to this property, the quantum state associated with a given spatial configuration will have a nonzero overlap with another spatial configuration. We show how this property can account for both semi-classical dynamics and unitary time evolution. We introduce an explicit toy model that follows semi-classical time evolution at the coarse-grained level; in addition, its time evolution is unitary at the fine-grained level. The only price that one must pay is the violation of the entropy bound; we discuss reasons why this may be justifiable.
\end{abstract}

\maketitle

\newpage

\tableofcontents

\newpage

\section{Introduction}

\begin{figure*}[t]
\includegraphics[width=\linewidth]{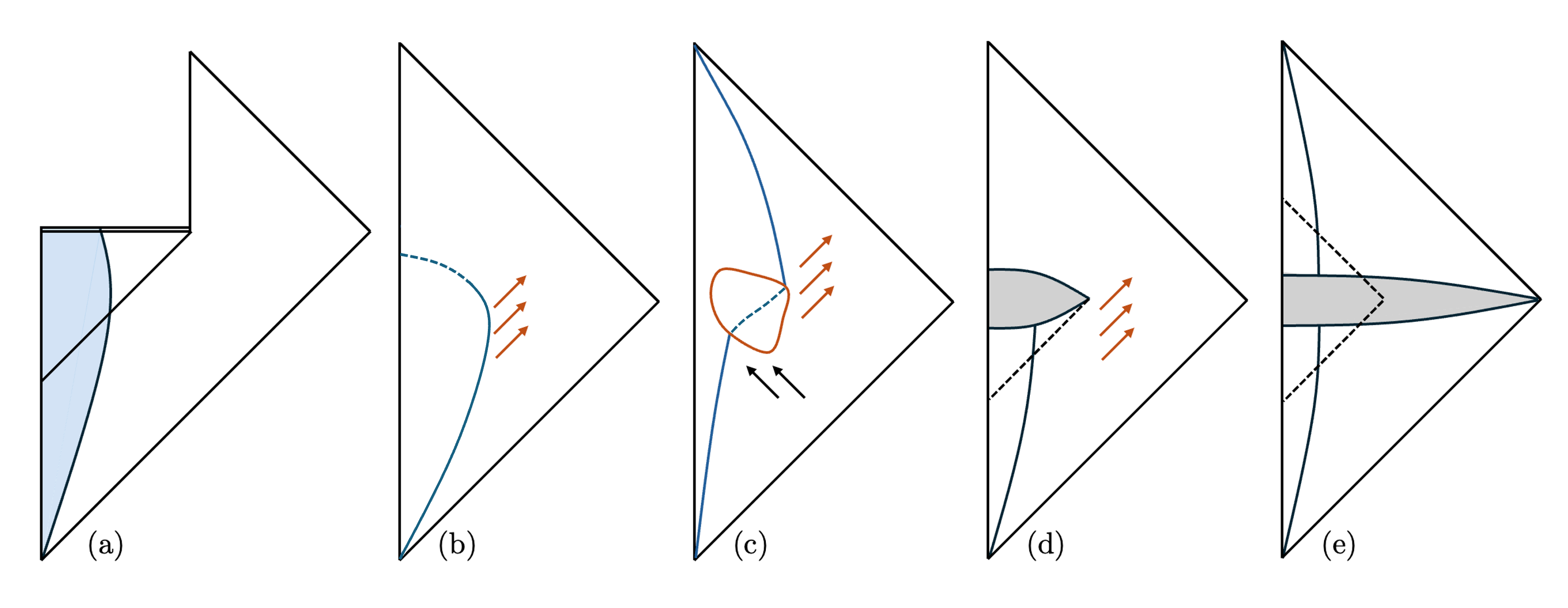}
\caption{\label{fig:1} (a) Penrose diagram for an evaporating black hole described by semi-classical physics. (b) The pre-Hawking radiation scenario describes a collapsing shell that emits radiation (red arrows) without forming an apparent horizon or singularity; this requires a spacelike bending part of the shell (dashed curve). (c) The regular black hole scenario has a circle-shaped apparent horizon (red curve), where the blue-colored curve is the core of the regular black hole, and this core part should be spacelike if it is trapped by the apparent horizon due to the collapsed matter (black arrows). (d) The Ashtekar-Bojowald scenario connects the black hole to Minkowski spacetime by resolving the spacelike singularity (gray-colored region). (e) The Haggard-Rovelli scenario connects the black hole phase to the white hole phase, requiring quantum gravitational effects that extend even outside the event horizon.}
\end{figure*}

The black hole information loss paradox \cite{Hawking:1976ra} remains one of the most difficult unresolved problems in modern theoretical physics. The paradox is grounded in and reveals the tension between well-established assumptions. On the one hand, the semi-classical causal structure of general relativity, combined with local quantum field theory, predicts the process of black hole formation and evaporation \cite{Hawking:1975vcx} (Fig.~\ref{fig:1}(a)). On the other hand, standard quantum mechanics stipulates that a closed system evolves from a pure state to another pure state; this implies that there is no loss or cloning of information \cite{Susskind:1993if}. The problem is that a black hole behaves like a thermal system, similar to a blackbody \cite{Bekenstein:1973ur}. If its thermodynamic entropy is the same as the statistical entropy~\cite{Strominger:1996sh}, then we expect the black hole to emit information, even if the black hole is large enough, and hence, Hawking radiation should carry information \cite{Page:1993wv}. The question is how information from inside the event horizon is transmitted to the outside, for which non-local \cite{Page:2013mqa} or radical effects near the horizon \cite{Almheiri:2012rt} in the sub-Planckian regime have been proposed, as well as the cloning of information \cite{Susskind:1993mu}. 

In this paper, we start from non-perturbative quantum gravity to understand black hole evaporation. While there are numerous approaches to non-perturbative quantum gravity, the most conservative, and the one considered the starting point for others such as loop quantum gravity, is known as \textit{canonical quantum gravity} \cite{DeWitt:1967yk}. This approach begins with the wavefunction of the universe~$\Psi$, which is a functional of the field configuration on a 3-hypersurface. This functional equation of $\Psi$ is a kind of Schr\"{o}dinger equation, known as the \textit{Wheeler-DeWitt equation}.

A central issue with the Wheeler-DeWitt equation is that it does not contain an explicit time parameter; hence, it is not possible to speak of unitary time evolution (of black holes). One proposal to bypass this problem is to introduce a clock using the Page-Wootters formalism~\cite{Page:1983uc}. However, introducing a relational time parameter is not sufficient; one must also introduce a series of time slices. The question remains of which quantum state to assign to a time slice. One possibility is to impose the property such that a hypersurface of a classical background must be a superposition of eigenstates (here, we can obtain eigenstates after we perform separation of variables of the Wheeler-DeWitt equation) and form a state, so to speak, a \textit{coherent state}. Intuitively speaking, a coherent state is a quantum state that most closely resembles a classical state insofar as it is a state of minimum uncertainty. The coherent state postulate was first proposed in \cite{Wang:2025siq}. A key feature of coherent states is that any two states are not fully distinguishable; if one expands the Hamiltonian in this basis of coherent states, one finds that there will be non-vanishing transition probabilities from one state to another.

In this paper, we embed this picture in scenarios without an event horizon, by which we mean the boundary of the past of the future null infinity. If there is no event horizon, a complete Cauchy surface will likely evolve consistently, and there must be no loss of information. However, we can ask what the information flow is in this background. Our Hamiltonian formulation can be generically applied to various scenarios without event horizons and provides a constructive explanation of information flow, for example, in terms of entanglement entropy. We do not claim that this is a final resolution of the information loss paradox; nevertheless, it provides a consistent and constructive way to explicitly show a flow of information.

Our paper is organized as follows. In Sec.~\ref{sec:pre}, we summarize previously proposed ideas on the scenarios without an event horizon. In Sec.~\ref{sec:sem}, we discuss the semi-classical dynamics according to the superposition of the spacetime proposal and the Hamiltonian dynamics at a coarse-grained level. In Sec.~\ref{sec:uni}, we extend this computation to the fine-grained level to demonstrate a unitary time evolution. Finally, in Sec.~\ref{sec:con}, we summarize this paper and discuss possible future topics for the ultimate understanding of the information loss paradox.

\section{\label{sec:pre}Preliminaries: black hole evaporation without an event horizon}

In this section, we revisit several proposed candidate ideas for the information-loss paradox, namely, scenarios without event horizons. We will identify the common features of these scenarios and, finally, embed them into the Hamiltonian formulation.

\subsection{Black holes without horizons}\label{sec:IIA}

First, we consider the scenario in which evaporating black holes never form a singularity or an event horizon. Fig.~\ref{fig:1} illustrates four scenarios that have been proposed previously. 

\begin{enumerate}
\item \textit{The pre-Hawking radiation scenario} considers the case where a collapsing shell emits radiation (red arrows) without forming an apparent horizon or singularity \cite{Baccetti:2016lsb}. Suppose we insist on this until the evaporation is complete. One obtains a semi-classical evaporating spacetime without a singularity, at the cost of introducing a region where the matter shell bends in a spacelike direction \cite{Chen:2017pkl}. This might be acausal, suggesting that unknown quantum gravitational effects are required to explain it. 
\item \textit{The regular black hole scenario} includes the formation of apparent horizons, but the singularity is avoided due to specific preparation of matter at the core (blue curve and blue dashed curve) \cite{Ayon-Beato:1999kuh}. This requires that the boundary of the core be timelike in an untrapped region, but spacelike in a trapped region. Hence, there must be a transition from a timelike to a spacelike direction in the core that might be acausal; this leads one to ask how quantum-gravitational effects explain this phenomenon.
\item \textit{The Ashtekar-Bojowald scenario} \cite{Ashtekar:2005cj} connects the black hole phase to the Minkowski phase in a causally connected patch, thanks to (hypothetical) quantum-gravitational effects. As one resolves the spacelike singularity (gray-colored region), one can extend spacetime beyond the region, and the putative event horizon (dashed line) has no causal meaning (because there is no singularity at all). However, there must be a way to connect spacetime slices before and after the evaporation, while loop quantum gravity predicts the local bouncing nature at the singularity \cite{Ashtekar:2018cay}. If quantum bouncing occurs at the spacelike singularity, one is led to ask how to embed the bouncing process within a causally connected patch.
\item \textit{The Haggard-Rovelli scenario} \cite{Haggard:2014rza} connects the black hole phase to the white hole phase to show that the quantum bouncing behavior is embedded in a single causally connected patch. However, this requires that quantum effects extend even beyond the horizon \cite{Brahma:2018cgr}. If this is the case, how can we justify these kinds of quantum effects from quantum gravitational principles?
\end{enumerate}

These four scenarios share a common point: \textit{we require quantum gravitational effects in a specific way for a particular spacelike surface}; for example, an acausal movement of matter from a timelike to spacelike direction for (b) and (c), or quantum gravitational effects for a spacelike hypersurface for (d) and (e). If one justifies the quantum effects for a spacelike domain, then one can ask whether we can embed the entire dynamics within a single causal patch or not. There are several proposals to resolve the singularity based on arguments in loop quantum gravity \cite{Ashtekar:2018cay}; however, it remains unclear whether there is a constructive mechanism to connect one spacelike hypersurface to another that extends beyond the horizon. In this regard, the Euclidean path integral approach provides a reasonable framework for such a mechanism.

\subsection{Euclidean path integral approach}

The Wheeler-DeWitt equation is the quantum Hamiltonian constraint equation \cite{DeWitt:1967yk}:
\begin{eqnarray}
\hat{\mathcal{H}} \Psi = 0,
\end{eqnarray}
where $\Psi \equiv \Psi[h_{ab}, \phi]$ is a functional of $3$-metric $h_{ab}$ and matter fields $\phi$. The formal solution of the Wheeler-DeWitt equation can be given via the path integral: the Euclidean path integral is a propagator from an in-state $| h^{i}_{ab}, \phi^{i} \rangle$ to an out-state $| h^{f}_{ab}, \phi^{f} \rangle$ \cite{Hartle:1983ai}:
\begin{eqnarray}
\langle h^{f}_{ab}, \phi^{f} | h^{i}_{ab}, \phi^{i} \rangle = \int_{i \rightarrow f} \mathcal{D} g_{\mu\nu} \mathcal{D}\phi \; e^{- S_{\mathrm{E}}},
\label{eq2euclidean}
\end{eqnarray}
where $S_{\mathrm{E}}$ is the Euclidean action and the path integral sums over all regular Euclidean geometries connecting the in-state to the out-state.

One advantage of the Euclidean approach is that there is a straightforward method for approximating the path integral, namely the steepest-descent approximation. Eq.~\eqref{eq2euclidean} is well approximated by Euclidean on-shell solutions that connect the in-state to the out-state, the so-called instantons. In particular,
\begin{eqnarray}
\langle h^{f}_{ab}, \phi^{f} | h^{i}_{ab}, \phi^{i} \rangle \simeq \sum_{i \rightarrow f} e^{- S^{\mathrm{on-shell}}_{\mathrm{E}}}.
\end{eqnarray}

Regarding the Euclidean instantons, we can summarize significant results as follows:
\begin{itemize}
\item There exists an instanton that connects a black hole to a Minkowski geometry \cite{Chen:2018aij} that is essential to explain the unitary evolution \cite{Maldacena:2001kr}. For a given entropy $S$, the transition probability is $\simeq e^{-S}$ \cite{Gregory:2013hja,Fischler:1990pk,Chen:2017suz}.
\item The tunneling probability to Minkowski dominates at late time \cite{Chen:2022eim}.
\end{itemize}

\begin{figure*}[t]
    \centering
    \includegraphics[width=0.75\linewidth]{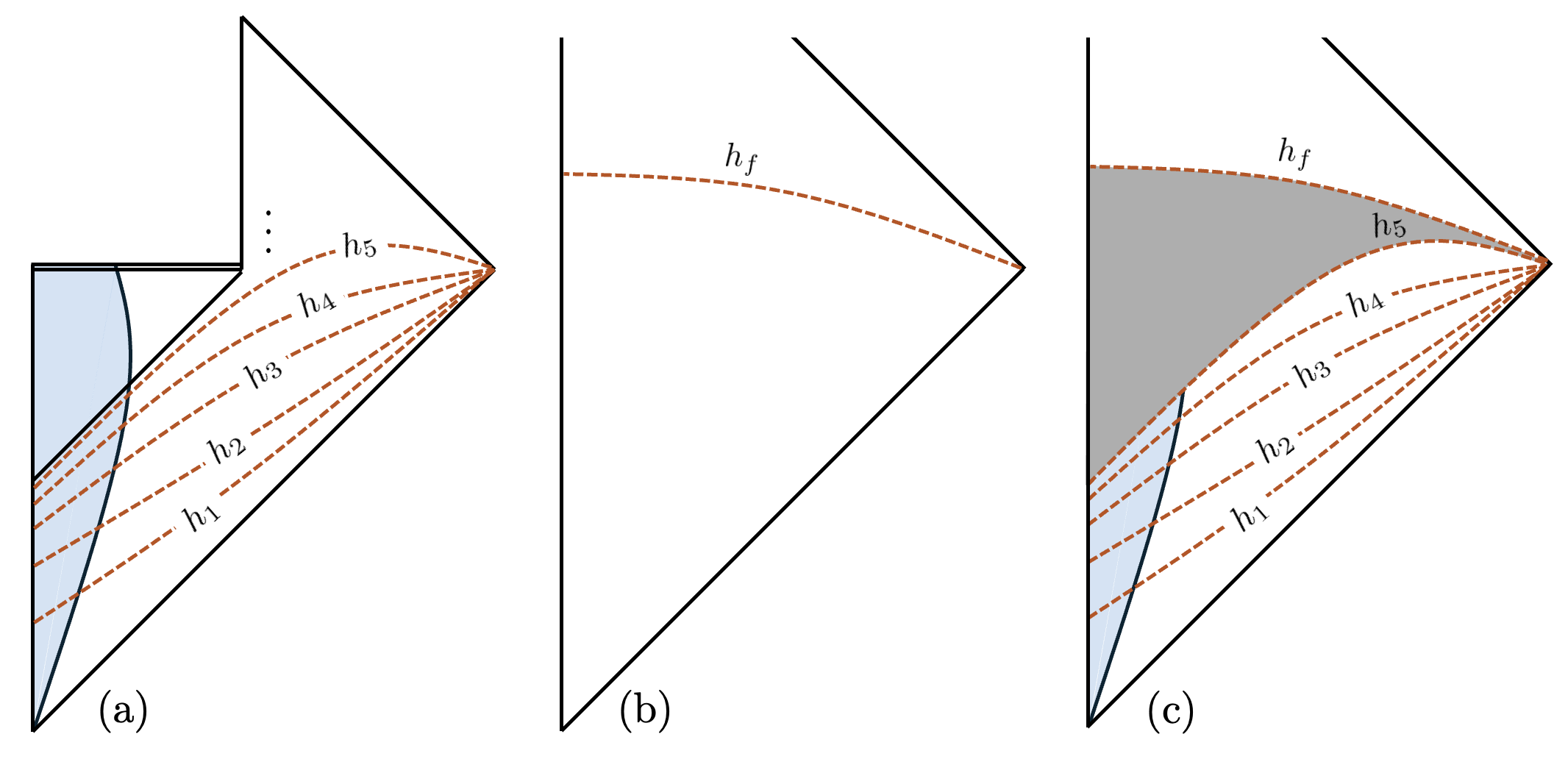}
    \caption{(a)-(b) Topology change will be mediated by the Euclidean propagator that connects from $| h_{i} \rangle$ ($i = 1, 2, ...$) to $| h_{f} \rangle$. To define the Euclidean propagation, we choose time slices outside the event horizon. (c) The approximate causal structure of the evaporating black hole according to the Euclidean path integral approach. The Euclidean propagator describes the gray-colored region.}
    \label{fig:2}
\end{figure*}

Tunneling from $h_{i}$ to $h_{f}$ is always possible, but it is exponentially suppressed at early times. Therefore, if we follow the dominant histories, at early times the semi-classical slices shown in Fig.~\ref{fig:2}(a) dominate (i.e., tunnelling from $h_{1}$ to $h_{5}$); but at the critical time when the probability of the trivial geometry $p_{2}$ is higher than that of the black hole geometry $p_{1}$, i.e., $p_{2} > p_{1}$, say at $h_{5}$, the spacetime dominantly transits to $h_{f}$, Fig.~\ref{fig:2}(b), corresponding to Minkowski spacetime filled by emitted radiation. Smoothly connecting $h_{5}$ to $h_{f}$ gives Fig.~\ref{fig:2}(c). The gray-colored region in Fig.~\ref{fig:2}(c) that smoothly connects from $h_{5}$ to $h_{f}$ is formally described by the Euclidean propagator. As a result, Fig.~\ref{fig:2}(c) is the dominant causal structure according to the Euclidean path integral \cite{Chen:2022eim}.

This picture effectively removes the event horizon and the singularity and was anticipated by Hawking himself \cite{Hawking:2014tga}, such that the contribution of the topologically trivial geometry would connect the collapsing phase to the entire evaporation process, ultimately removing the event horizon.

\subsection{Summary}

In the previous subsection, we summarized several proposed approaches for removing the event horizon. A common feature of these proposals is the requirement of effects beyond semi-classical gravity on a spacelike hypersurface; we need the help of quantum gravity for either an acausal bending of the matter field or a transition between spacelike surfaces that connect the before and after of evaporation. Supposing that quantum-gravitational phenomena can account for such effects, one can provide a closed causal structure with no event horizon or singularity. On top of this causal structure, one can choose a series of time slices that cover only the region outside the event horizon. Of course, there exists a fully quantum-gravitational domain, but except for that part, we can reasonably guess that the time slices are sufficiently semi-classical. Hence, the quantum gravitational wave function for these time slices should be described as coherent states.

All models agree on the conclusion: \textit{there is no information loss if there is no event horizon}. Answering whether this causal structure preserves information is the goal of the remainder of this work. To understand the flow of information, we will adopt Fig.~\ref{fig:2}(c) as the universal causal structure for models with no event horizon. The important physical assumptions are as follows:
\begin{enumerate}
\item The causal structure itself is justified independently based on an approach to quantum gravity, justifying the spacelike transition region in a way.
\item The time slices, except for the spacelike quantum gravitational regime, correspond to coherent states (for a recent discussion, see \cite{Lin:2025jmz}).
\end{enumerate}
Assuming these two conditions, we can apply the same analysis presented in this paper, even when focusing specifically on the model in Fig.~\ref{fig:2}.

\section{\label{sec:sem}Semi-classical evolution in the coarse-grained description}

To investigate the unitary time evolution of a black hole spacetime, let us first introduce a toy model that describes the semi-classical evolution at the coarse-grained level.

\subsection{The functional Schr\"{o}dinger equation}

To introduce time dependence into the Wheeler-DeWitt equation, we introduce an ideal clock, characterized by Hamiltonian $\hat{\mathcal{H}}_C$ and covariant time observable $\hat T$, satisfying the canonical commutation relation $[\hat T, \hat{\mathcal{H}}_C] = i \hslash$. Eigenstates of $\hat T$ are the time states $| t \rangle$ (describing the instant of time $t$ read by the clock), translations of which are generated by $\hat{\mathcal{H}}_C$.  {For simplicity, we work with the scenario where the clock and gravitational sectors are decoupled. We}
write the Wheeler-deWitt equation in the form 
\begin{eqnarray}
\hat{\mathcal{H}} \Psi = \big( \hat{\mathcal{H}}_G+\hat{\mathcal{H}}_C \big) \Psi =0,
\end{eqnarray}
where $\hat{\mathcal{H}}_G$ is the Hamiltonian for the gravitational sector and $\hat{\mathcal{H}}_C$ is the Hamiltonian for the clock sector.  The condition of decoupled clock and gravitational sectors restricts the model, but there nevertheless exist important physical settings in which this condition is satisfied, such as homogeneous cosmologies with a massless scalar field \cite{Hoehn:2019fsy}. For our model, we imagine that all possible spatial configurations are asymptotically flat and assume the clock to be situated ``at infinity.'' The clock may then be arbitrarily close to ideal, and we approximate it as such. For a general cosmology, no single clock may furnish a global time \cite{Kuchar_Time_Interpretation}. We assume, then, that any dynamics described by our model are contained within a local patch for which the clock is valid.

We further assume that the dynamics of the clock and gravitational degrees of freedom are correlated,
\begin{eqnarray}
| \Psi \rangle = \int dt \: | t \rangle | \psi (t) \rangle,
\end{eqnarray}
where $| \psi (t) \rangle$ satisfies the functional Schr\"{o}dinger equation  (as per Page and Wootters) \cite{Page:1983uc,Vachaspati:2006ki}:
\begin{eqnarray}
i \frac{\partial}{\partial t} | \psi \rangle = \hat{\mathcal{H}}_G | \psi \rangle
\end{eqnarray}
and may be interpreted as the state of the gravitational sector conditioned upon a reading of the clock in some state $| t \rangle$. For black holes, the functional Schr\"{o}dinger equation of the gravitational sector can describe the time evolution of the quantum state with respect to this faraway clock and is manifestly unitary. Despite this, it remains to be shown whether this unitary evolution can lead to the complete evaporation of the black hole. The other question is: by what mechanism can the information carried by Hawking radiation be recovered at the end of this process?

Regarding the complete evaporation of the black hole, we require, for a black hole represented by the quantum state $| n \rangle$, that:
\begin{enumerate}
\item \textit{Nonorthogonality of quasi-classical states}: We postulate that coherent states describe quasi-classical metrics of time slices. In addition, the quantum states of time slices are not orthogonal to each other, so to speak,
\begin{eqnarray}
\langle \tilde{m} | \tilde{n} \rangle \simeq \varepsilon^{|\tilde{n} - \tilde{m}|},\label{eq:as1}
\end{eqnarray}
where $\varepsilon \ll 1$ is a small number.
\item \textit{Energy}: The quantum Hamiltonian of the gravitational degrees of freedom can be obtained by quantizing the ADM energy $E$, i.e., 
\begin{eqnarray}
    E \rightarrow \hat{\mathcal{H}}_G = \sum_n E_n | n \rangle\langle n | 
\end{eqnarray}
with $\langle n | m \rangle = \delta_{nm}$. If we adopt the postulate that quasi-classical metrics are described by coherent states, then we propose making the replacement $| n \rangle \to | \tilde{n} \rangle$, where $\langle \tilde{n} | \tilde{m} \rangle \simeq \varepsilon^{| \tilde{n} - \tilde{m} |}$. Hence, the number $\tilde{n}$ represents the energy $E_{\tilde{n}}$ of the state $| \tilde{n} \rangle$ and the Hamiltonian is approximately written by
\begin{eqnarray}
\hat{\mathcal{H}}_G \simeq \sum_{\tilde{n}} E_{\tilde{n}} | \tilde{n} \rangle \langle \tilde{n} |.\label{eq:as2}
\end{eqnarray}
\end{enumerate}
Assuming these two conditions, the steepest-descent of the wave function will evolve along the contour satisfying
\begin{eqnarray}
\frac{dE_{\tilde{n}}}{dt} \propto \Delta E_{\tilde{n}}.
\end{eqnarray}
Therefore, by choosing $E_{\tilde{n}}$ appropriately, one can recover the semi-classical dynamics of an evaporating black hole via the functional Schr\"{o}dinger equation. Indeed, it has been shown~\cite{Wang:2025siq} that by assuming the Stefan-Boltzmann law, one recovers the semi-classical evaporation law $M_\mathrm{BH}(t) \sim -t^{1/3} + \mathrm{const}$.

\subsection{Toy model of evaporating black holes}

Let us now introduce a toy model of an evaporating black hole to gain insight into the mechanism by which information may be retrieved during evaporation. Our goal is to capture the essential physical aspects of unitary black hole evaporation and to connect them to the prior postulate of the nonorthogonality of quasi-classical metrics.

\subsubsection{Single-particle state}

We consider a particle labeled by $i$. Let us define the state of the particle $i$, where
\begin{eqnarray}
| 1 \rangle_{i} =
\begin{pmatrix}
    1 \\
    0 \\
    0
\end{pmatrix}, \;\;\;\;
| 0 \rangle_{i} = A
\begin{pmatrix}
    \varepsilon \\
    1 \\
    \varepsilon
\end{pmatrix}, \;\;\;
| -1 \rangle_{i} =
\begin{pmatrix}
    0 \\
    0 \\
    1
    \end{pmatrix}.
\end{eqnarray}
Here, $A = 1/ \sqrt{1+2\varepsilon^{2}}$ is a normalization constant. We heuristically refer to each state as a particle $(|1\rangle)$, a vacuum $(|0\rangle)$, and an antiparticle $(|-1\rangle)$, respectively.

We consider the following Hamiltonian (represented by an orthogonal basis):
\begin{eqnarray}
\hat{H}_{i} \equiv \omega
\begin{pmatrix}
    1 & 0 & 0\\
    0 & 0 & 0\\
    0 & 0 & -1
\end{pmatrix}.
\end{eqnarray}
Because of this choice of the Hamiltonian, the expectation of the energy of the particle state is $+\omega$, the antiparticle state is $-\omega$, and the vacuum state is $0$. In other words,
\begin{align}
_{i}\langle j |\hat{H}_{i} | j \rangle_{i} &= j \omega \quad \mathrm{for} \quad j = 0, \pm 1 
\end{align}
and
\begin{eqnarray}
_{i}\langle 1 | 0 \rangle_{i} =\, _{i}\langle 0 | -1 \rangle_{i} \simeq \varepsilon.
\end{eqnarray}
Hence, we specify three states (particle, vacuum, antiparticle), and the particle and antiparticle states overlap with the vacuum state. This can mimic the process by which vacuum fluctuations can create a particle state with positive energy and an antiparticle state with negative energy.

\subsubsection{Multi-particle state}

\begin{figure}[h]
\begin{center}
\includegraphics[scale=0.5]{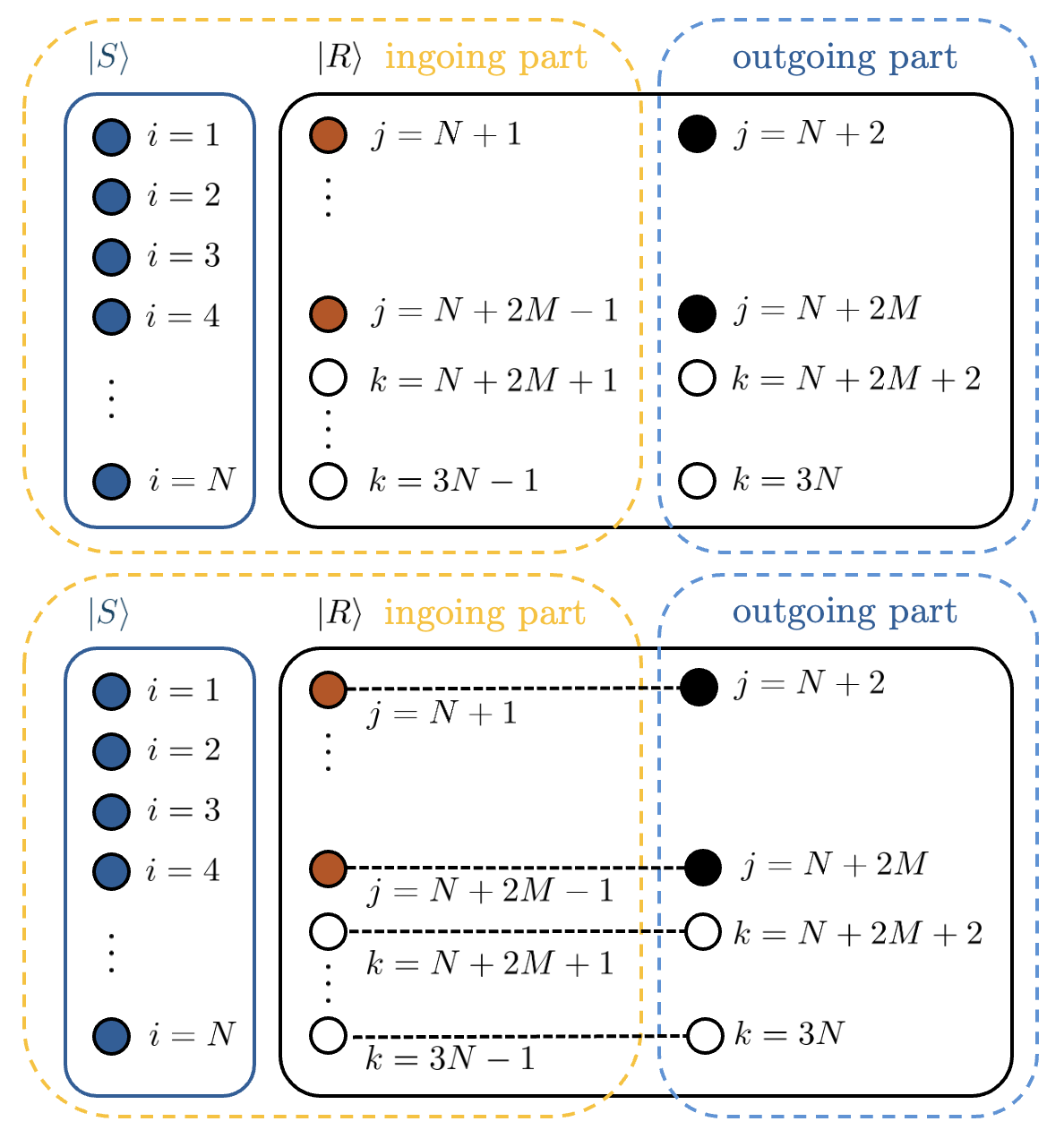}
\caption{\label{fig:3} (Top panel) Particle indices, where $| S \rangle$ is the interior state of the star and $| R \rangle$ is the Hawking radiation state. Blue- and black-colored dots are particle states $| 1 \rangle$, red-colored dots are antiparticle states $| -1 \rangle$, and white-colored dots are vacuum states $| 0 \rangle$. The yellow-colored box is for the incoming part, and the blue-colored box is for the outgoing part. (Bottom panel) One can introduce entanglement, where the yellow-colored dotted lines denote maximal entanglement between two particles.}
\end{center}
\end{figure}

Now we consider a system with $3N$ particles--see Fig.~\ref{fig:3}(a). We partition the system into an \textit{interior} subsystem:
\begin{eqnarray}
| S \,\rangle = \bigotimes_{i = 1}^{N} | 1 \rangle_{i}.
\end{eqnarray}
which models the interior of the collapsing matter, say a star, 
and the \textit{Hawking radiation} subsystem:
\begin{align}
| R \,\rangle &= \left( \bigotimes_{j = 1}^{M} | -1 \rangle_{N+2j-1} \otimes | 1 \rangle_{N+2j} \right) \otimes \left( \bigotimes_{k = 1}^{N-M} | 0 \rangle_{N+2M+2k-1} \otimes | 0 \rangle_{N+2M+2k} \right),
\end{align}
where $M < N$. This heuristically means there are $M$ particle-antiparticle pairs and $N-M$ zero-point fluctuations.

The joint state of the star and Hawking radiation is thus $| \psi \rangle = | S \rangle \otimes | R \rangle$.  Here, the point is that we can uniquely specify $| \psi \rangle$ by $M$ or $\tilde{n} \equiv N - M$, and hence we can henceforth denote $| \psi \rangle \equiv | \tilde n \rangle$ for $\tilde n = 0, 1, 2 , \hdots , N$. It is straightforward to verify that $\langle \tilde{m} | \tilde{n} \rangle \simeq \varepsilon^{2|\tilde{n} - \tilde{m}|}$.

\begin{figure*}[t]
\includegraphics[width=\linewidth]{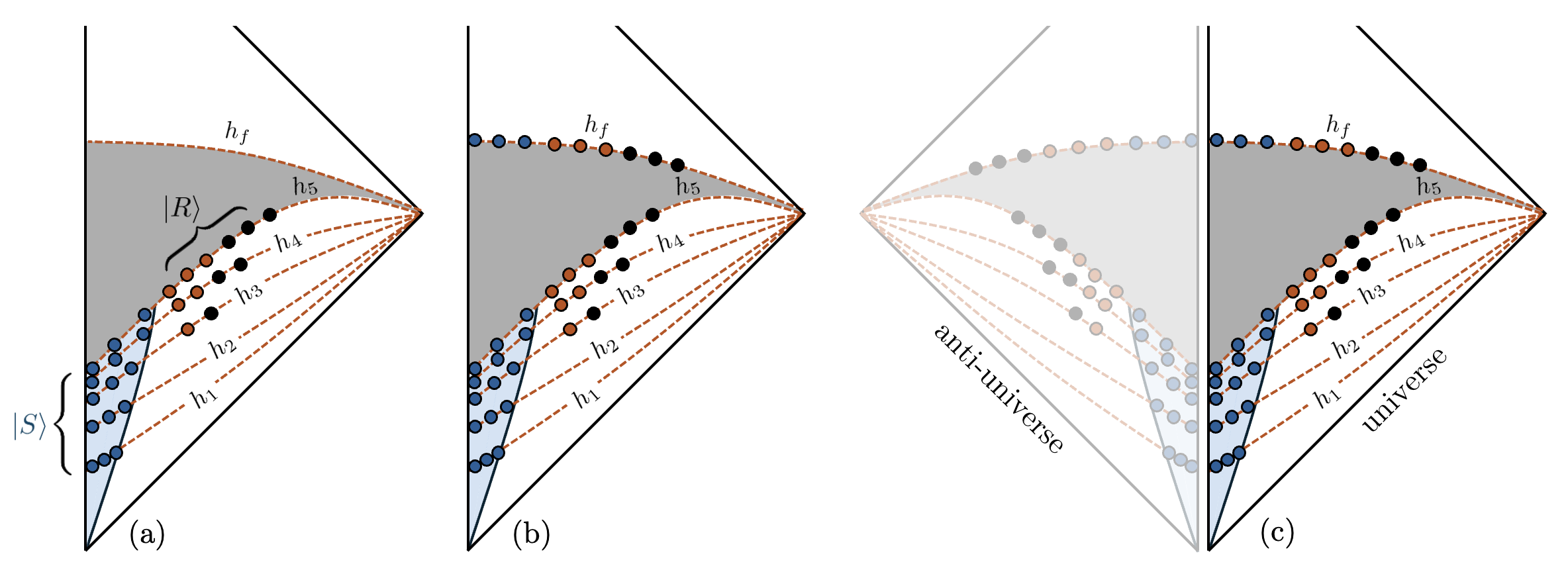}
\caption{(a) A toy model of an evaporating black hole. Here, we did not denote vacuum fluctuations. As time goes on, the number of particle-antiparticle pairs increases as $M$ increases or $\tilde{n}$ decreases. (b) According to semi-classical computations, the black hole should totally evaporate, and all quantum states propagate to $h_{f}$. (c) A universe and an anti-universe, where each particle state in the anti-universe has exactly opposite energy expectation values.}
\label{fig:4}
\end{figure*}

We next construct the Hamiltonian. Defining $\hat{\mathcal{H}}_i \equiv I^{\otimes (i-1)} \otimes \hat H_i \otimes I^{\otimes (3N-i)}$,   the total Hamiltonian is then
\begin{eqnarray}
\hat{\mathcal{H}} \equiv \sum_{i=1}^{3N} \hat{\mathcal{H}}_{i}.
\end{eqnarray}
The total energy is $\langle \tilde n | \hat{\mathcal{H}} | \tilde n \rangle = N \omega$, which is independent of~$\tilde{n}$. 
However, we can define the energy of the incoming part:
\begin{eqnarray}
E_{\tilde{n}} = \left\langle \sum_{i=1}^{N} \hat{\mathcal{H}}_{i} + \sum_{j=1}^{M} \hat{\mathcal{H}}_{N+2j-1} \right\rangle = \tilde{n} \omega,
\label{eq19}
\end{eqnarray}
where the expectation value is taken with respect to $| \tilde n \rangle$. This process can be drawn on top of the causal structure, as shown in Fig.\ \ref{fig:4}(a)-(b).

From this figure, as $M$ increases, the number of particle-antiparticle pairs increases. If we focus only on the ingoing part, as $M$ increases, the energy decreases, while the total energy remains constant. Therefore, if $M$ can be related to the time variable, this toy model can explain a continuous accumulation of particle-antiparticle pair creations, although the total energy is always conserved. We now ask whether we can justify that $M$ decreases over time.

\subsubsection{Semi-classical evolution with coarse-grained states}

We can ask whether we can be sure that the quantum states will evolve toward evaporation as shown in Fig.~\ref{fig:4}(a)-(b). To understand this, we need to solve the time-dependent Schr\"{o}dinger equation.

Now, let us define a reduced Hamiltonian
\begin{eqnarray}
\hat{\mathcal{H}}' \equiv \sum_{i=1}^{N} \hat{\mathcal{H}}_{i} + \sum_{j=1}^{M} \hat{\mathcal{H}}_{N+2j-1}, 
\end{eqnarray}
such that Eq.~(\ref{eq19}) can be rewritten in the following form,
\begin{eqnarray}
\hat{\mathcal{H}}' = \sum_{\tilde{n} = 1}^{N} E_{\tilde{n}} | \tilde{n} \rangle \langle \tilde{n} |,
\end{eqnarray}
where $E_{\tilde{n}} = \tilde{n} \omega$. Of course, this reduced Hamiltonian is not for the global quantum state, but we can trace out the outgoing particles $| \tilde{n} \rangle \equiv | \tilde{n}' \rangle \otimes | N - \tilde{n}' \rangle \rightarrow | \tilde{n}' \rangle$ where
\begin{eqnarray}
| \tilde{n}' \rangle \equiv | S \rangle \bigotimes_{j = 1}^{M} | -1 \rangle_{N+j} \bigotimes_{k = 1}^{N-M} | 0 \rangle_{N+M+k},
\end{eqnarray}
$| N - \tilde{n}' \rangle$ is the complement part of $| \tilde{n}' \rangle$, and
\begin{align}
    \hat{\mathcal{H}} &\equiv \hat{\mathcal{H}}' \otimes I_{N} + I_{N} \otimes \hat{\mathcal{H}}'',
\end{align} 
with 
\begin{align} 
    \hat{\mathcal{H}}' \equiv \sum_{i=1}^{2N} \hat{\mathcal{H}}_{i} &\simeq \sum_{\tilde{n}' = 1}^{N} E_{\tilde{n}'} | \tilde{n}' \rangle \langle \tilde{n}' |,
    \\
    \hat{\mathcal{H}}'' \equiv \sum_{i=2N+1}^{3N} \hat{\mathcal{H}}_{i} &\simeq \sum_{\tilde{n}' = 1}^{N} \left( N\omega -E_{\tilde{n}} \right) | N - \tilde{n}' \rangle \langle N - \tilde{n}' |.
\end{align}
Here, the last two equations are approximately accurate. This $| \tilde{n}' \rangle$ state is a mimicker of a black hole (even if we choose the hypersurfaces outside the horizon) for the ingoing part, and the $| N - \tilde{n}' \rangle$ state is for the outgoing part.  Note that we have not yet introduced entanglement into the model.   Henceforth  we call the state $| \tilde{n} \rangle$ as a \textit{fine-grained} state, while $|\tilde{n}' \rangle$ as a \textit{coarse-grained} state.

Finally, the following equation will be (approximately) satisfied:
\begin{eqnarray}\label{eq:fs}
i \frac{\partial}{\partial t} | \psi' \rangle \simeq \hat{\mathcal{H}'} | \psi' \rangle,
\end{eqnarray}
where
\begin{eqnarray}
| \psi' \rangle \equiv \sum_{\tilde{n}'} a_{\tilde{n}'}(t) | \tilde{n}' \rangle
\end{eqnarray}
and $a_{\tilde{n}'}(t)$ are normalization constants. Based on our arguments, we can now demonstrate that there is a time evolution in the $\tilde{n}$-decreasing direction.

\subsubsection{Boundary conditions}

One interesting property of the functional Schr{\"o}dinger equation, Eq.~(\ref{eq:fs}), is that the energy of the state $| \psi' \rangle$ must be conserved. On the other hand, in the context of the information loss problem, we expect the black hole's energy to decrease over time. To reconcile these two points of view, we point out a clue in the boundary condition. Indeed, the functional Schr{\"o}dinger equation is a first-order differential equation of the time parameter (here parametrized by the reading of a clock decoupled from the gravitational degrees of freedom). Hence, we have no freedom to provide the initial time derivative of a wave packet. As a result, if there is an energy-decreasing branch of the wave packet, there should also exist an alternative energy-increasing branch. 

Let us suppose the existence of the energy-increasing branch in an alternative universe, a so-called \textit{anti-universe} that has exactly the opposite energy states of the universe. We illustrate this in Fig.\ \ref{fig:4}(c). One way to achieve this is to postulate a global quantum state $| \Psi \rangle$
\begin{eqnarray}
| \Psi \rangle = | \bar{\psi} \rangle \otimes |\psi \rangle,
\end{eqnarray}
where $| \bar{\psi} \rangle = | - \tilde n \rangle \equiv | \bar S \rangle \otimes | \bar R \rangle$ is the quantum state of the anti-universe, where
\begin{align}
    | \bar{S} \,\rangle &\equiv \bigotimes_{i = 1}^{N} | -1 \rangle_{i}, \\
    | \bar{R} \,\rangle &\equiv \left( \bigotimes_{j = 1}^{M} | 1  \rangle_{N+2j-1}  \otimes | -1 \rangle_{N+2j} \right) \otimes  
     \left( \bigotimes_{k = 1}^{N-M} | 0 \rangle_{N+2M+2k-1} \otimes | 0 \rangle_{N+2M+2k} \right)
\end{align} 
and 
\begin{align} 
    \hat{\mathcal{H}}_{\mathrm{tot}} &\equiv \hat{\mathcal{H}} \otimes I_{3N} + I_{3N} \otimes \hat{\mathcal{H}} \nonumber 
    \\
    &= \left( \hat{\mathcal{H}}' \otimes I_{N} + I_{2N} \otimes \hat{\mathcal{H}}'' \right) \otimes I_{3N} + I_{3N} \otimes \left( \hat{\mathcal{H}}' \otimes I_{N} + I_{2N} \otimes \hat{\mathcal{H}}'' \right)
    \nonumber \\
    &= \left( \hat{\mathcal{H}}' \otimes I_{N} \otimes I_{3N} + I_{3N} \otimes \hat{\mathcal{H}}' \otimes I_{N} \right) + \left( I_{2N} \otimes \hat{\mathcal{H}}'' \otimes I_{3N}  + I_{3N} \otimes I_{2N} \otimes \hat{\mathcal{H}}'' \right) 
    \nonumber \\
    &\equiv \hat{\mathcal{H}}_{\mathrm{tot}}' \otimes I_{N} + I_{N} \otimes \hat{\mathcal{H}}_{\mathrm{tot}}'',
    \label{eq30}
\end{align}
where in the last line, the Hamiltonian is written in the $| \pm \tilde{n}' \rangle$ and $| \pm N \mp \tilde{n}' \rangle$ basis. This setup implies that each particle state in the anti-universe has exactly the opposite energy expectation value to that in the original universe, Fig.~\ref{fig:4}(c). Hence $\langle \Psi | \hat{\mathcal{H}}_\mathrm{tot} | \Psi \rangle = 0$; from the global quantum state $| \Psi \rangle$, one can conclude that the total energy of each slice (including the universe and the anti-universe) is zero, which is analogous to \textit{Dirac's sea}.

\begin{figure}[h]
    \centering
    \includegraphics[scale=0.75]{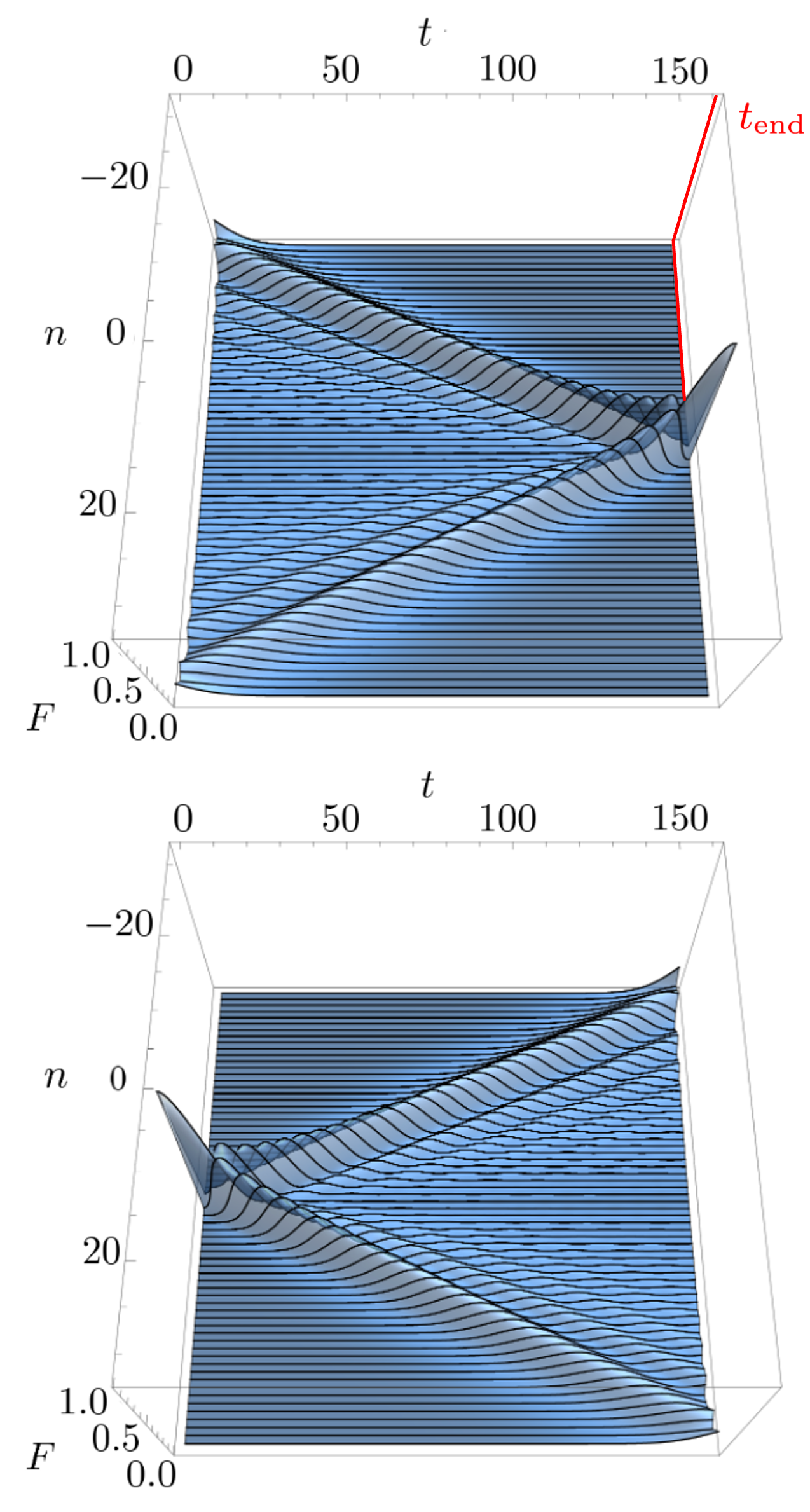}
    \caption{(top) The dynamics of the quantum state of the ingoing part satisfying the boundary conditions Eq.~(\ref{eq:b1}). Two waves cross at the last time $t = t_{\mathrm{end}}$ and reach their maximum at $n = 0$; this implies that the black hole should be completely evaporated at the end. The lower and upper wave packets indicate the ingoing part of the universe and the anti-universe, respectively. (bottom) The dynamics of the quantum state of the outgoing part satisfying the boundary conditions Eq.\ (\ref{eq:b4}). Two waves cross at the initial time $t = 0$ and reach their maximum at $n = 0$. The lower and upper wave packets indicate the outgoing part of the universe and the anti-universe, respectively.}
    \label{fig:inside-evolution}
\end{figure}

At the coarse-grained level, the total Hamiltonian can be decomposed by the black hole mimicker state $| \tilde{n}' \rangle$ and its anti-universe state $| -\tilde{n}' \rangle$:
\begin{equation}
\hat{\mathcal{H}}'_{\mathrm{tot}} \simeq \sum_{\tilde{n}} \left( - \tilde{n} \omega \right) | - \tilde{n}' \rangle \langle -\tilde{n}' | \otimes I + I \otimes \sum_{\tilde{n}} \left( \tilde{n} \omega \right) | \tilde{n}' \rangle \langle \tilde{n}' |,
\end{equation}
where $\tilde{n} = 0, 1, ... , \infty$. Since the sum of the energy of $| - \tilde{n}' \rangle$ and $| \tilde{n}' \rangle$ is zero, it is reasonable to impose a boundary condition on the wave packet at the final time, so to speak, $t = t_{\mathrm{end}}$:
\begin{eqnarray}
a_{\tilde{n}'}( t_{\mathrm{end}}) = \bar a_{\tilde n'} ( t_{\mathrm{end}} ) =  \delta_{\tilde{n}'0}, \label{eq:b1} 
\end{eqnarray}
where
\begin{eqnarray}
| \psi' \rangle &\equiv& \sum_{\tilde{n}'} a_{\tilde{n}'}(t) | \tilde{n}' \rangle, \\
| \bar{\psi}' \rangle &\equiv& \sum_{\tilde{n}'} \bar{a}_{\tilde{n}'}(t) | - \tilde{n}' \rangle,
\end{eqnarray}
and the global coarse-grained state is $| \bar{\psi}' \rangle \otimes |\psi' \rangle$. 
This boundary condition implies that the black hole has evaporated completely at $t_{\mathrm{end}}$. If we evolve the solution backward in time, there are two branches: one that increases energy (backward in time) and another that decreases energy (backward in time). In other words, if we move forward in time, there are two solutions, one of which is the decreasing-energy wave packet in our universe. In contrast, the other is the increasing-energy wave packet in an anti-universe. Two wavepackets annihilate at $t = t_{\mathrm{end}}$, and this guarantees complete evaporation; see Fig.~\ref{fig:inside-evolution} top panel.

In addition, one can obtain the Hamiltonian for the outgoing part:
\begin{align}
    \hat{\mathcal{H}}''_{\mathrm{tot}} \simeq \sum_{\tilde{n}} \left( -N + \tilde{n} \right) \omega | -N + \tilde{n}' \rangle \langle -N +\tilde{n}' | \otimes I  + I \otimes  \sum_{\tilde{n}} \left( N- \tilde{n} \right) \omega | N -\tilde{n}' \rangle \langle N -\tilde{n}' | .
\end{align}
One may alternatively provide the boundary condition of the wave packet of the outgoing part at the initial time, so to speak, $t = 0$:
\begin{eqnarray}
b_{\tilde{n}'}( 0 ) &= 
\bar{b}_{\tilde{n}'}( 0 ) = \delta_{\tilde{n}'0}, \label{eq:b4}
\end{eqnarray}
where
\begin{eqnarray}
| \psi'' \rangle &\equiv& \sum_{\tilde{n}'} b_{\tilde{n}'}(t) | N - \tilde{n}' \rangle, \\
| \bar{\psi}'' \rangle &\equiv& \sum_{\tilde{n}'} \bar{b}_{\tilde{n}'}(t) | -N + \tilde{n}' \rangle,
\end{eqnarray}
and the global coarse-grained state is $| \bar{\psi}'' \rangle \otimes |\psi'' \rangle$. Therefore, as time goes on, the energy of the outgoing part monotonically increases, while the energy of its counterpart in the anti-universe monotonically decreases; see Fig.~\ref{fig:inside-evolution} bottom panel.

To summarize the physically relevant points of our construction, including the anti-universe:
\begin{itemize}
\item The oppositely behaving solution branch (to that of our universe) is always located within an anti-universe; as a result, it guarantees complete evaporation. Also, the energy conservation of each coarse-grained sector is ensured.
\item If there is no interaction between two universes, we assume that the quantum states of the universe and anti-universe are always a product. 
\item As one includes the outgoing part (and with a suitable boundary condition of the outgoing part), conservation of energy in each universe is guaranteed, even though the ingoing energy decreases and the outgoing energy increases.
\end{itemize}
Hence, it is not surprising that we focus only on the part of our universe that does not violate unitarity or energy conservation. It is worth noting that the existence of the antiuniverse is a natural consequence of unitarity and energy conservation, analogous to the Dirac sea.

Finally, we locate this process on top of Fig.~\ref{fig:4}(a)-(b). Initially, at $h_{1}$, we define the star interior state $|S\rangle$. Particles in this state evolve in a time-like direction (blue-colored region and blue dots). As time goes on, Hawking particle pairs are created, say after $h_{3}$. From our coarse-grained computations, we can be sure that the number of Hawking particles should monotonically increase, say for $h_{4}$ and $h_{5}$. The black dots are outgoing, while the blue and red dots are ingoing. Hence, the black hole mass (the energy of the blue dots plus the red dots) should monotonically decrease.  Eventually, around the endpoint of the evaporation (around $h_{5}$), the slice should transit to the topologically trivial geometry $h_{f}$, and all particles (star interior, antiparticles, and particles) should appear at $h_{f}$, Fig.~\ref{fig:4}(b).

\subsubsection{Thermodynamic interpretation}

We can now ask whether the toy model is consistently related to the semi-classical computations of black hole evaporation. Let us consider a static black hole. For given $N$, the energy is
\begin{eqnarray}
E = \omega N
\end{eqnarray}
and the entropy is (if we assume that $| 1 \rangle$ and $| -1 \rangle$ are the only physical degrees of freedom)
\begin{eqnarray}
S = N \log 2.
\end{eqnarray}
Therefore,
\begin{eqnarray}
\frac{dS}{dE} = \frac{1}{T} = \frac{\log2}{\omega},
\end{eqnarray}
and hence, $T$ is a constant.

For $1+1$-dimensional spacetime, the Stefan-Boltzmann law imposes that
\begin{eqnarray}
\frac{dE}{dt} = - \alpha T^{2} \propto \mathrm{const.},
\end{eqnarray}
where $\alpha$ is a constant. Hence, if we regard $dt \propto \Delta n$, then
\begin{eqnarray}
\left| \frac{dE}{dt} \right| \propto \left| \frac{\Delta E_{\tilde{n}}}{\Delta \tilde{n}} \right| = \omega \propto \mathrm{const.},
\end{eqnarray}
which is qualitatively consistent. As we know that this system will totally evaporate according to the Stefan-Boltzmann law, it is not surprising that the Schr\"{o}dinger equation allows the same wave propagation according to the semi-classical evolution.

\subsection{General remarks}

As we conclude this section, let us pause to make some key remarks about our framework. 
\begin{enumerate}
\item To introduce time-dependence into the Wheeler-DeWitt equation, we have adopted a time-slicing corresponding to a clock at asymptotic infinity, subject to certain caveats. Thus, we consider only slices outside the horizon; hence, there are two components of the joint state of our universe—the star's interior and the Hawking radiation.
\item The Hawking radiation state $| R \rangle$ contains pairs of positive and negative energy particles, initially in a separable state. Usually, these two particles are entangled. We will introduce the entanglement and its implications in the next section.
\item We have introduced the zero-energy states by hand. We assume that the zero-energy states provide the non-orthogonality condition to particle or antiparticle states. This allows the possibility that the vacuum state creates a particle-antiparticle pair. Hence, the vacuum fluctuations trigger the evaporation.
\item We assume that the energy of every quantum is a constant $\omega$. However, in general, $\omega$ can be a continuous variable. This might be refined by introducing more detailed assumptions to ensure consistency with the Stefan-Boltzmann law in $(3+1)$ dimensions.
\end{enumerate}

\section{\label{sec:uni}Unitary evolution in the fine-grained description}

In the previous section, we demonstrated that our black hole toy model satisfies the assumptions of semi-classical dynamics (Eqs.~(\ref{eq:as1}) and (\ref{eq:as2})). We have separated the quantum states into three parts: the star interior, ingoing antiparticles, and outgoing particles. We have defined the ingoing part as the star interior plus the ingoing antiparticles. The coarse-grained state, or the quantum state of the ingoing parts, which encapsulates the star's interior and ingoing radiative degrees of freedom, describes semi-classical evolution consistent with black hole thermodynamics. In this section, we extend this description to the total state to verify the unitarity of its time evolution. We consider a fine-grained description that includes both ingoing and outgoing parts of quantum states and ask whether it remains unitary after evaporation. In particular, we will focus on the behavior of the entanglement entropy under the assumption that the particle-antiparticle pairs are maximally entangled. 

\subsection{Introducing entanglement}

First, let us assign an additional spin-1/2 degree of freedom to each particle. We use the nomenclature
\begin{eqnarray}
| 1, j \rangle_{i} \equiv | 1
\rangle_{i} \otimes | j  \rangle_{i} \quad \mathrm{for} \: j = \:\uparrow, \downarrow.
\end{eqnarray}
Using this notation, we define the quantum states of the star interior
\begin{eqnarray}
| S \,\rangle = \bigotimes_{i = 1}^{N} | 1, \uparrow \rangle_{i}.
\end{eqnarray}
Here, for simplicity, we assume that all interior microstates have spin-up.
In reality, the quantum state would be a statistical ensemble; this does not change our main result. The quantum state of the Hawking radiation becomes
\begin{align}
| R \,\rangle &= \bigotimes_{j = 1}^{M} \frac{1}{\sqrt{2}} \left( | -1, \uparrow \rangle_{N+2j-1} \otimes | 1, \downarrow \rangle_{N+2j} + | -1, \downarrow \rangle_{N+2j-1} \otimes | 1, \uparrow \rangle_{N+2j} \right) \nonumber \\
& \bigotimes_{k = 1}^{N-M} \frac{1}{\sqrt{2}}\left( | 0, \uparrow \rangle_{N+2M+2k-1} \otimes | 0, \downarrow \rangle_{N+2M+2k} + | 0, \downarrow \rangle_{N+2M+2k-1} \otimes | 0, \uparrow \rangle_{N+2M+2k} \right).
\end{align}
The total state $| \psi \rangle$ is defined as
\begin{eqnarray}
| \psi \rangle = | S \rangle \otimes | R \rangle.
\end{eqnarray}
The bottom panel of Fig.~\ref{fig:3} illustrates the inclusion of entanglement in $| R \rangle$. As long as we do not change the Hamiltonian $\mathcal{\hat{H}}$, the energy expectation value and several previous relationships are not changed. However, we can compute non-trivial entanglement entropy.

\subsection{Three entropies: entanglement, Bekenstein-Hawking, and Boltzmann}

\begin{figure}[h]
    \centering
    \includegraphics[scale=0.4]{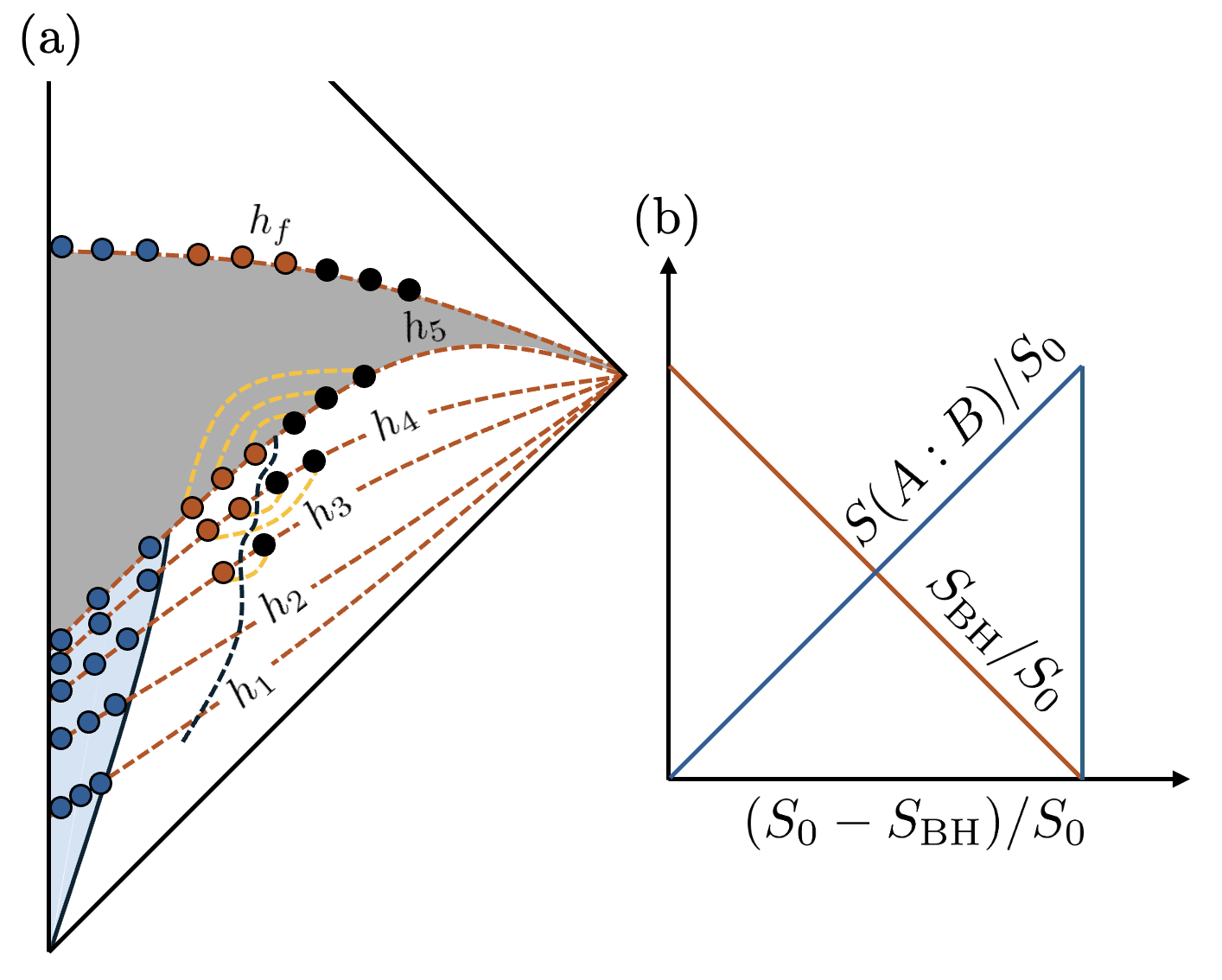}
    \caption{(a) A conceptual picture of the boundary (black-colored dashed curve), where its left side is the ingoing part $A$, and the other side is the outgoing part $B$. (b) A conceptual modified Page curve, where the red-colored curve is the Bekenstein-Hawking entropy and the blue-colored curve is the entanglement entropy of the radiation.}
    \label{fig:5}
\end{figure}

We have $3N$ particles. We reiterate to separate the system such that the $i = 1$, ..., $N$-th particles (star interior) and the $i = N+1$, $N+3$, ..., $N + 2M - 1$-th particles (antiparticles) are defined as the \textit{ingoing part} $A$, while their remnant (particles and vacuum fluctuations) is defined as the \textit{outgoing part} $B$. Since all entangled particles are maximally entangled, the entanglement entropy between $A$ and $B$ for a given $M$ is
\begin{eqnarray}
S(A : B) = M \log 2.
\end{eqnarray}
As the black hole evaporates, the entanglement entropy monotonically increases, Fig.~\ref{fig:5}(a).
On the other hand, in the coarse-grained description of the black hole state, $| \tilde{n}' \rangle$, there are $N$ particles and $M$ antiparticles (with $N-M$ zero-point fluctuations, which we will not count because entanglement among vacuum states is not physically meaningful). Therefore, the Bekenstein-Hawking entropy (apparent Boltzmann entropy of the black hole) is
\begin{eqnarray}
S_{\mathrm{BH}} = \left( N - M \right) \log 2,
\end{eqnarray}
because the black hole is represented by its energy, which is proportional to $N-M$. One important point is that our time slice is outside the event horizon, and hence the actual thermodynamic entropy might be much smaller than this value. So, we can regard $S_{\mathrm{BH}}$ as the maximum thermodynamic entropy of a given configuration.

Therefore, at the Page time, $M = N/2$, and the Bekenstein-Hawking entropy bound must be violated ($S(A:B) > S_{\mathrm{BH}}$). On the other hand, if we regard the true degrees of freedom, then there are $N + M$ particles, and therefore, the true Boltzmann entropy is
\begin{eqnarray}
S_{\mathrm{B}} = \left( N + M \right) \log 2.
\end{eqnarray}
Of course, $S_{\mathrm{B}} > S(A:B)$ is always satisfied.

\subsection{Modified Page curve and unitary evolution}

The implication of the previous subsections is that in our toy model, the entanglement entropy increases monotonically until the end of evaporation. At the endpoint, there is no more separation between ingoing and outgoing parts, and hence, the entanglement entropy drops to zero, Fig.~\ref{fig:5}(b). The behavior shown in Fig.~\ref{fig:5}(b) describes unitary evolution; however, at some point the entanglement entropy exceeds the Bekenstein-Hawking entropy. Hence, the entropy bound relation must break down. We will discuss the implications of this in the Appendix. Similar behavior was observed in Euclidean path integral computations \cite{Chen:2022eim}. Likewise, one can compute the entanglement entropy as follows:
\begin{eqnarray}
\langle S \rangle = p_{1} S_{1} + p_{2} S_{2},
\end{eqnarray}
where we denote the subscript $1$ as the history with a black hole (collapsed matter with apparent horizon) with probability $p_{1}$, the subscript $2$ as the history without an event horizon with probability $p_{2}$, $S_{1} = S_{0} - S$ is the entanglement entropy of $1$, which is a monotonically increasing function, while $S_{2} = 0$, since the black hole and radiation are located in the same space because there is no event horizon. Here, $S_{0}$ is the initial Bekenstein-Hawking entropy, while $S$ is the Bekenstein-Hawking entropy at a certain moment. The point is that initially $p_{1} \gg p_{2}$, but, eventually, $p_{2} \gg p_{1}$ will be realized, keeping $p_{1} + p_{2} = 1$. Hence, although the entanglement entropy $S_{1}$ increases monotonically at early times, the quantity $\langle S \rangle$ will increase to a maximum and then
eventually   drop to zero. One can then realize a modified Page curve with a smooth transition to zero entropy (in contrast with the abrupt change in Fig.~\ref{fig:5}(b)). 

One may ask when such a transition happens. It is a probabilistic process, but one dominated by the case in which the black hole loses almost all of its energy to outgoing radiation (this is guaranteed by our formalism, e.g., Fig.~\ref{fig:inside-evolution}); that is the reason why we denote, in Fig.~\ref{fig:5}(a), that the Euclidean transition happens when the number of ingoing antiparticles is the same as the number of particles in the star. In reality, there might be a lot more particle-antiparticle pairs than particles in the star. The generalization of our description to more realistic cases will be left for future work.

\begin{figure*}[t]
    \centering
    \includegraphics[width=0.725\linewidth]{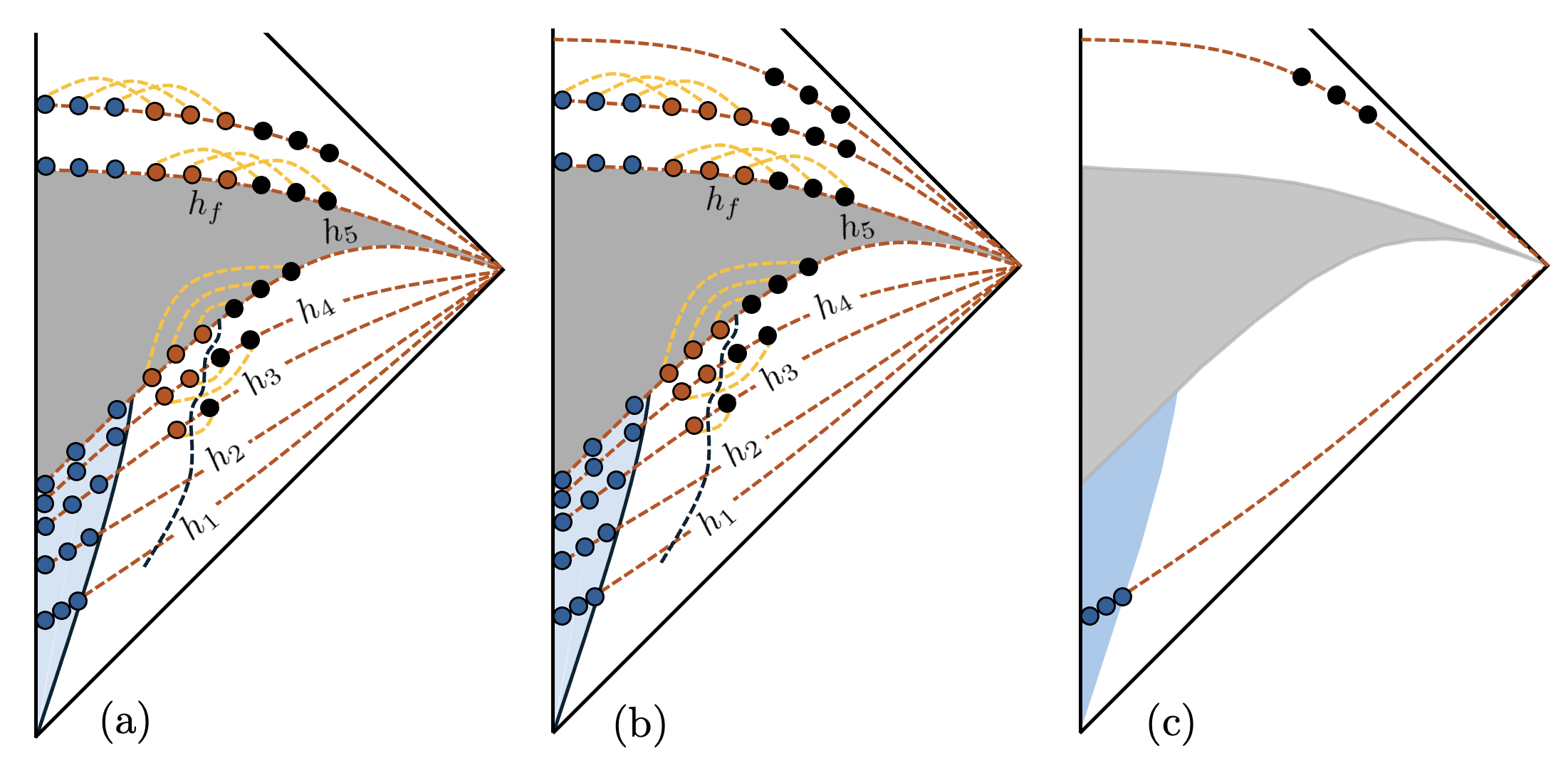}
    \caption{(a) After the total evaporation, introducing unitary operations, one can transfer entanglement to the star interior particles and antiparticles of Hawking radiation. (b) Since these entangled particles are zero-energy separable combinations, they can be traced out. (c) Finally, we obtain a unitary evolution from a collapsing star to outgoing radiation.}
    \label{fig:11}
\end{figure*}

\subsection{State reductions by unitary operations}

After the evaporation finishes, we will obtain the state
\begin{align}
    | R \,\rangle &= \bigotimes_{j = 1}^{N} \frac{1}{\sqrt{2}} ( | -1, \uparrow \rangle_{N+2j-1} \otimes | 1, \downarrow \rangle_{N+2j} + | -1, \downarrow \rangle_{N+2j-1} \otimes | 1, \uparrow \rangle_{N+2j} ).
\vphantom{\bigotimes_{i=1}^N}
\end{align}
Now we consider the unitary operation (e.g., a swapping operation that randomizes the system) on the state $| \psi \rangle$, Fig.~\ref{fig:8}(a). Finally, we aim to obtain the state with the following form, Fig.~\ref{fig:11}(b):
\begin{align}
    | \psi \rangle_{f} &= | A \rangle \otimes | B \rangle 
    \vphantom{\bigotimes_{i=1}^N}
    \\
    &\equiv \bigotimes_{i=1}^{N} \frac{1}{\sqrt{2}} ( | 1,  \uparrow \rangle_{i} \otimes | -1, \downarrow \rangle_{N+2i-1} + |1, \downarrow \rangle_{i} \otimes | -1, \uparrow \rangle_{N+2i-1} )  \otimes | B \rangle,
\end{align}
where $|B \rangle$ is the state of $N$ particles of the outgoing part. Note that $| A \rangle$ is separable with zero total energy. Hence, we can unitarily trace out $ | A \rangle$, and as a result, we will obtain $ N$ particles in the outgoing radiation, Fig.~\ref{fig:11}(c). Note that, for this protocol, we do not need the non-orthogonal state $| 0 \rangle$ anymore. Everything can work by operating on orthogonal states.

\begin{figure*}[t]
\includegraphics[width=0.825\linewidth]{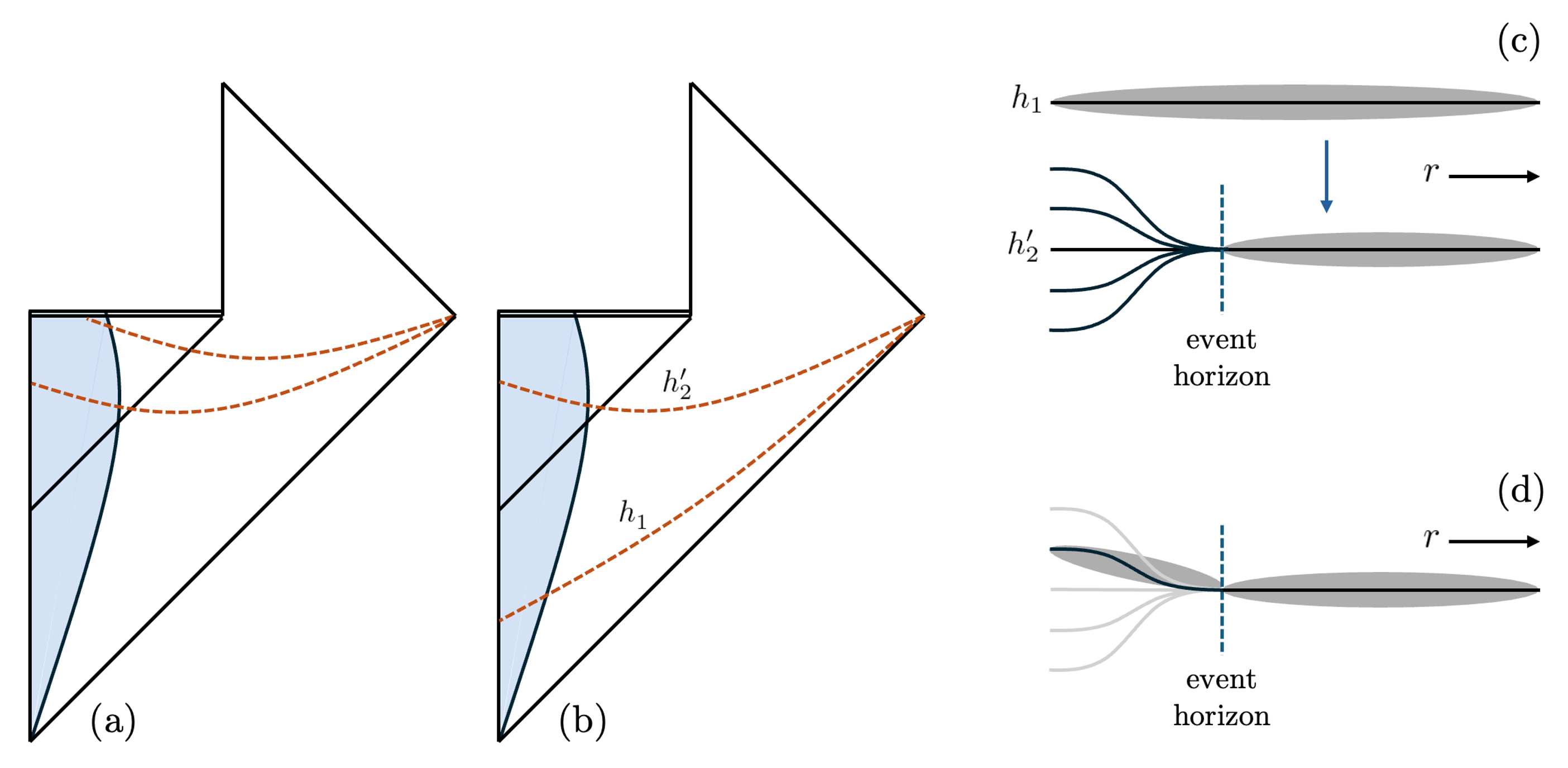}
\caption{To discuss the infall problem, let us revive the notion of the event horizon. (a) Regarding the infall problem, can we choose the time slice that crosses the event horizon or even touches the singularity? (b) One may assume the initial slice at $h_{1}$ will evolve into $h_{2}'$, which crosses the event horizon. (c) The classical background of $h_{1}$ (black line) is a coherent state in terms of the wave function (gray-colored region represents). We hypothesize that $h_{1}$ will evolve to $h_{2}'$, where inside the event horizon is a superposition of different classical backgrounds and loses the nature of a coherent state, while outside the event horizon is still a coherent state. (d) Hence, if we observe a specific classical geometry inside the event horizon, one needs a new clock and already loses unitarity.
}
\label{fig:9}
\end{figure*}

\section{\label{sec:con}Conclusion}

\subsection{Summary of the scenario}

In this paper, we provide a constructive perspective on the information loss paradox. Our approach was inspired by the framework of canonical quantum gravity, which is based on a Hamiltonian for the matter and geometry degrees of freedom. Even though the wavefunction(al) of the universe is static, by adopting the Page-Wootters formalism and conditioning on the time reading of an asymptotic clock, we obtain a functional Schr\"{o}dinger equation describing unitary evolution.

We assume that our clock and hypersurface choice is sufficiently nice in the sense that the quantum state of the hypersurface is a good coherent state. As a result, an overlap exists between two coherent states, allowing one to transition to the other. The transition process from one state to another likewise follows from the Sch{\"o}dinger evolution and is consistent with the semi-classical description on a coarse-grained level.

While everything in our construction is explicitly unitary, one consequence is that the entanglement entropy can exceed its Boltzmann entropy, i.e., the entropy bound relation may be violated. However, we provide several examples in the Appendix that indicate why this might be a reasonable price to pay. Of course, after evaporation, there might be a unitary operation that reduces the number of states and eventually recovers the entropy bound.

Our scenario shares some commonalities with previous candidate resolutions. The strength of our description lies in our reliance on the well-defined first principle of canonical quantum gravity. Additionally, our description applies to a wide range of candidate scenarios without an event horizon, as discussed in Sec.~\ref{sec:IIA}. Moreover, our scenario will circumvent the pathologies of black hole complementarity by accepting the possibility of violations of the entropy bound.

However, we should note that our scenario is not complete unless we address the infall problem consistently. Although it is beyond the scope of this paper, we provide several comments in the next section.

\subsection{Infall problem: clock and hypersurface}

Now, let us ask what happens to an observer who falls into the black hole. What does the observer experience, and can this be described in the canonical formalism? This question can be well formulated in terms of the choice of the clock and the hypersurfaces. To describe this possibility, we briefly revisit the notion of the event horizon. Can we consistently represent the evolution of the wave function even if the hypersurface crosses over the event horizon or touches the singularity (see Fig.~\ref{fig:9})?

To address the question of unitarity, we need to introduce time into the Wheeler-DeWitt equation. To do this, we introduce the clock and obtain a functional Schr\"{o}dinger equation that has an explicit time dependence. As we introduce the explicit time dependence, we introduce a series of time slices across the entire spacetime. In principle, these slices can cross the event horizon. Then, the question regarding our approach is this: what prevents us from choosing time slices that cross the event horizon, or, conversely, what problems arise if we do?

To this end, we suggest the following. A given time slice presents a classical spacetime, and hence we expect that the corresponding wave function is well described as a coherent state. However, if a given putative time slice cannot be described as a coherent state, we cannot choose that classical slice to describe the unitary evolution. We are considering an evaporating black hole; our initial time slice was at past infinity. From past infinity, we smoothly deform the coherent wave packet to the next time step. If there is no consistent coherent state description for the interior of the event horizon, it is impossible to smoothly and continuously deform the time slices that evolve inside the event horizon. If this is not allowed, then to describe the state inside the horizon, we need to switch the clock. For example, one can choose a free-falling observer and use the observer's proper time observable as a clock. In this case, the free-falling clock is no longer related to the asymptotic observer's clock, and hence, there is no consistent and unitary description in terms of an asymptotic observer, even though the free-falling observer experiences a well-defined history.

To make the entire picture consistent, the asymptotic observer should exclude a free-falling clock. One may conclude that \textit{the black hole interior might not be a coherent state in terms of the unitary observer's clock}; rather, so to speak, the black hole interior might be a superposition of several coherent states from the perspective of the asymptotic observer, Fig.~\ref{fig:9}(b)-(c). On the other hand, an infalling observer's clock can describe a coherent state for the interior, Fig.~\ref{fig:9}(d); they are incommensurable. To justify this assertion, we need a more careful quantum gravitational analysis inside the horizon. This assertion might be supported by solving the Wheeler-DeWitt equation (see \cite{Lin:2025jmz}); we postpone this interesting project for future investigation.

\subsection{Generalization to field-theoretic approaches}

To establish a connection between the black hole evaporation process and the functional Schr\"{o}dinger equation, Eqs.~(\ref{eq:as1}) and (\ref{eq:as2}), we introduced an explicit toy model. This toy model provides an intuitive picture of the black hole evaporation process. However, it is fair to say that this is a toy model; hence, we need to extend the field-theoretical configurations. Recent work by Akil et.\ al.\ \cite{Akil:2025coj} takes a step in this direction, describing the Hawking particle pair by the state
\begin{eqnarray}
| R \rangle \propto \sum_{\omega} e^{- 4\pi M \omega} | - \omega \rangle \otimes | + \omega \rangle,
\end{eqnarray}
where $\omega$ is a continuous variable. One can justify this relation from a rather formal description of quantum field theory. Furthermore, by accumulating this Hawking radiation, one may recover the consistent time evolution of the wavefunction in accordance with the 3+1-dimensional Stefan-Boltzmann law. We leave this for future research, but it is reasonable to believe that unitary evolution is guaranteed even with this refinement.

\newpage

\section*{Acknowledgments}
The authors would like to thank Joshua Foo and Achintya Sajeendran for the important discussion at the earlier stage of this project. SW acknowledges support from the Natural Sciences and Engineering Research Council of Canada. DY was supported by the National Research Foundation of Korea (NRF) grant funded by the Korean government (No.~RS-2026-25476711).

\section*{Appendix. Meaning of the entropy bound violation}

Here, we consider what happens if the entropy bound is violated. We first illustrate several examples that demonstrate entropy bound violations; the entropy bound might be violated if we consider non-perturbative effects \cite{Mann:2025ojd} or an accumulation of perturbative effects \cite{Bae:2020lql}.

\subsubsection{Example 1: Isentropic process of a charged black hole}

Let us consider an isentropic process of a charged black hole with mass $M$ and charge $Q$ \cite{Mann:2025ojd}:
\begin{align}
    ds^{2} = - \left( 1 - \frac{2M}{r} + \frac{Q^{2}}{r^{2}} \right) dt^{2} + \left( 1 - \frac{2M}{r} 
    + \frac{Q^{2}}{r^{2}} \right)^{-1} d r^{2}  + r^{2} d\Omega_{2}^{2}.
\end{align}
By keeping the isentropic condition (the area or the horizon size $r_{+}$ is invariant), if one sends a particle with mass $m$ and charge $q$ from outside to inside the horizon, the following equation of motion is satisfied:
\begin{align}
\left( \frac{d r}{d \tau} \right)^{2} + V_{\mathrm{eff}}(r) &= 0 
\end{align} 
with the effective potential, 
\begin{align} 
V_{\mathrm{eff}}(r) &\equiv -\frac{1}{m^{2}} \left( E - q\frac{Q}{r} \right)^{2} + \left( 1 - \frac{2M}{r} + \frac{Q^{2}}{r^{2}} \right),
\end{align}
where $E$ is the asymptotic energy of the particle, and the isentropic condition is
\begin{eqnarray}
E - q\frac{Q}{r_{+}} = 0.
\end{eqnarray}
Therefore, there exists a finite sized potential barrier ($V_{\mathrm{eff}}$) for $r \geq r_{+}$. This classically prevents the isentropic process.

However, if we consider barrier penetration, the isentropic process can be allowed with the probability $\Gamma \simeq e^{-2S_E}$, where
\begin{align}
    S_{\mathrm{E}} &= \int_{r_{+}}^{r_{\mathrm{max}}} \frac{1}{\sqrt{V_{\mathrm{eff}}}(r)} dr,
\end{align}
where $V(r_{+}) = V(r_{\mathrm{max}}) = 0$ and $V(r) > 0$ for $r_{+} \leq r \leq r_{\mathrm{max}}$.

Therefore, if the black hole was initially maximally entangled, or if such a process accumulates, then the entanglement entropy can increase monotonically while the Bekenstein-Hawking entropy remains constant. Hence, at some moment, the entropy bound might be violated due to non-perturbative effects.

\subsubsection{Example 2: Non-thermal early Hawking radiation}

Even when considering perturbative effects, if they accumulate, the entropy bound can be violated. Let us consider early black hole radiation (before the Page time).

Suppose we have a bipartite system, where $A$ is the inside of the black hole with $m$ degrees of freedom and $B$ is the outside of the black hole with $n$ degrees of freedom; we assume that the total number of states, $n \times m$, is a constant. One can define the Boltzmann entropies $S_{A} \equiv \log m$ and $S_{B} \equiv \log n$, as well as the entanglement entropy $S(A:B)$. One can also define the distinguishable amount of information for $A$ and $B$ as follows: $I_{A} \equiv S_{A} - S(A:B)$ and $I_{B} \equiv S_{B} - S(B:A)$ \cite{Lloyd:1988cn}. This measures the difference from thermal equilibrium if $I_{A,B} > 0$. Finally, one can prove that the sum of the three information measures is constant for a pure state \cite{Alonso-Serrano:2017gis}:
\begin{eqnarray}
I_{A} + I(A:B) + I_{B} = \log mn =\mathrm{const.},
\end{eqnarray}
where $I(A:B)$ is the mutual information between $A$ and $B$.

Typical Hawking radiation involves the creation of a particle-antiparticle pair. There exists maximal entanglement between the incoming antiparticle and the outgoing particle. Hence, as the black hole evaporates, the entanglement entropy will increase maximally over time, at least before the Page time \cite{Page:1993wv}:
\begin{eqnarray}
\log n \simeq S(A:B).
\end{eqnarray}
Hence, in the early stage of black hole evaporation, Hawking radiation is approximately thermal (the von Neumann entropy is the same as the Boltzmann entropy) and will not carry distinguishable information. In addition to this, if we assume that the areal entropy corresponds to the Boltzmann entropy inside the black hole, we can write
\begin{eqnarray}
\log n = \frac{\mathcal{A}_{0} - \mathcal{A}}{4},
\end{eqnarray}
where $\mathcal{A}_{0}$ is the initial area of the black hole, and $\mathcal{A}$ is the area of the black hole at a certain time.

However, one can think of a situation where two Hawking pairs (so to speak, two antiparticles and two particles) can interact before the absorption of antiparticles \cite{Bae:2020lql}, for example, by introducing a $\lambda \phi^{4}$-term \cite{Leahy:1983vb}. This will result in a decrease in the entanglement between two particles and two antiparticles. If this process is cumulative then  
\begin{eqnarray}
S_{\mathrm{acc}} \equiv \log n - S(A:B) > 0
\end{eqnarray}
is satisfied. Then, there might be a non-thermal amount of information from the early stage of Hawking radiation
\begin{eqnarray}
I_{B} > 0.
\end{eqnarray}
Suppose this information originates from the interior of the black hole. In that case, an observer can witness the duplicated information \cite{Bae:2020lql}, and hence, this does not make sense.

What we can consistently interpret is that the increased amount of information $I_{B}$ is not from the black hole, but an artifact due to the interaction between Hawking pairs. Hence, due to this process, the total number of states $n \times m$ should increase, where the Boltzmann entropy inside the black hole is
\begin{eqnarray}
\log m = \log m_{0}n_{0} - \log n + S_{\mathrm{acc}},
\end{eqnarray}
where $m_{0}n_{0}$ is the initial number of states the total system. Here, $\log n = (\mathcal{A}_{0} - \mathcal{A})/4$ and $\log m_{0}n_{0} = \mathcal{A}_{0}/4$. Therefore,
\begin{eqnarray}
\log m > \frac{\mathcal{A}}{4}
\end{eqnarray}
is obtained. This shows that the entropy bound should be violated.

\subsubsection{Summary}

These two examples suggest that the entropy bound may be violated due to non-perturbative effects or the accumulation of perturbative effects. Therefore, we suggest that the violation of the entropy bound relation is not a serious matter in the context of evaporating black holes \cite{Buoninfante:2021ijy}.

\begin{figure}
    \centering
    \includegraphics[scale=0.3]{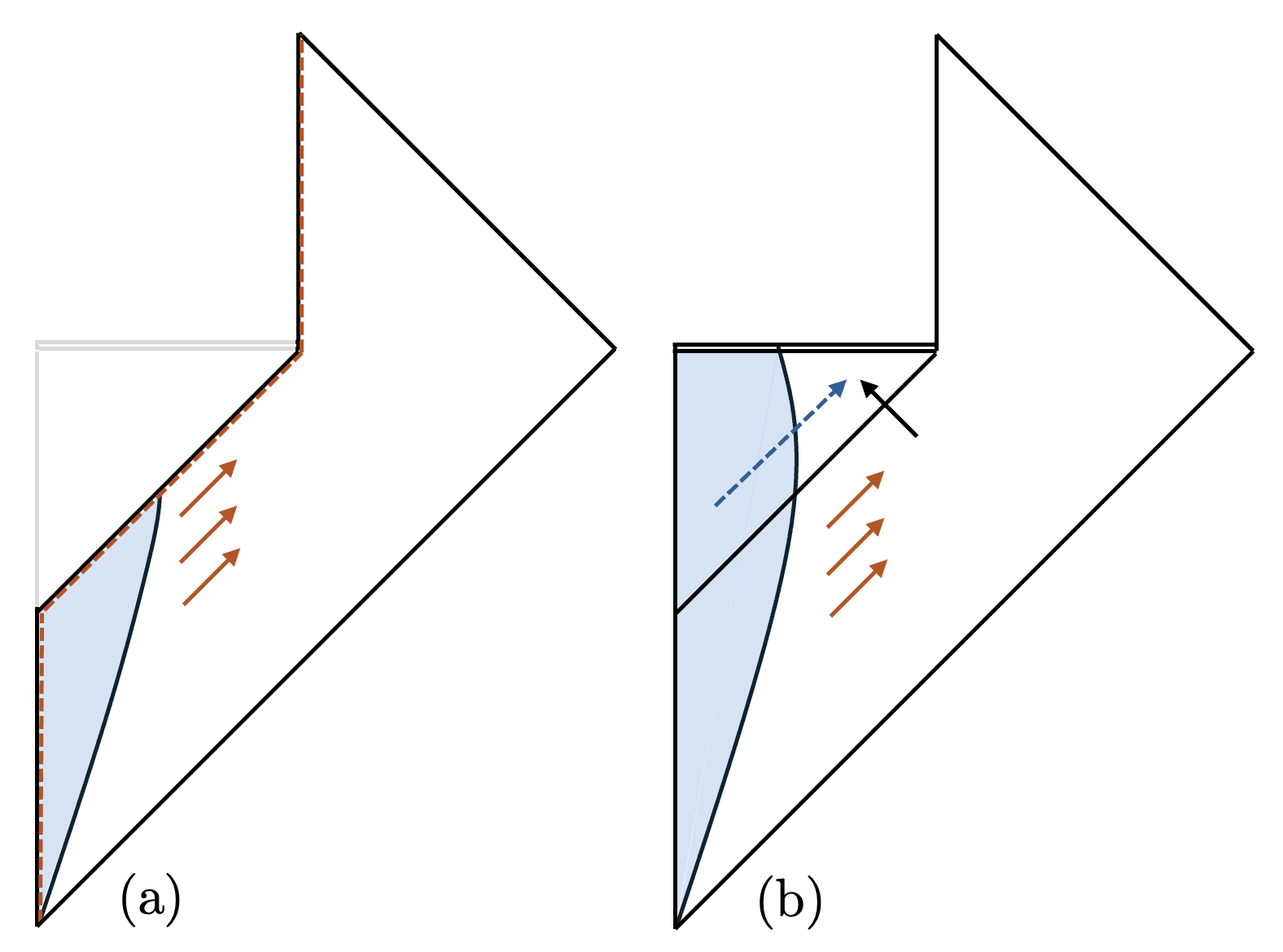}
    \caption{(a) The black hole complementarity scenario says that outside the stretched horizon (red dashed curve) is causally connected and time evolution is unitary in this patch. Hence, the collapsed matter is attached to the stretched horizon and eventually emitted by Hawking radiation without loss of information. (b) However, assuming that information escapes after the Page time by Hawking radiation, there should be a duplicated copy of information, where one is inside (propagating along a blue dashed arrow) and the other is outside (propagating along a red dashed arrow). There can be an observer (black arrow) who can witness the duplication of information.}
    \label{fig:8}
\end{figure}

Violating the entropy bound is essential to bypass the inconsistency argument of black hole complementarity. The original version of black hole complementarity says that there are two different descriptions of evaporating black holes, Fig.~\ref{fig:8}(a) \cite{Susskind:1993if}. One is for the asymptotic observer, and the other is for the infalling observer. For the asymptotic observer, information never crosses the event horizon; instead, information becomes attached to the stretched horizon (a timelike surface outside the event horizon, red dashed curve in Fig.~\ref{fig:8}(a), thermalized at the stretched horizon, and is emitted by Hawking radiation (red arrows); for the infalling observer, information is carried inside the event horizon until it touches the singularity. These two observers contradict each other because the initial information is duplicated, with one being inside and the other outside the horizon. Black hole complementarity holds that two observers are complementary because no observer can witness the inconsistency between their descriptions. However, this assertion is wrong in the sense that one can construct an observer who can see the duplicated information, Fig.~\ref{fig:8}(b) \cite{Yeom:2009zp}.

However, compared to Fig.~\ref{fig:2}(c), there exists a common point that \textit{the black hole evolution is unitary in terms of the asymptotic observer if we choose slices only outside the horizon}. Of course, the details differ, and our picture does not suffer from the duplication issue because we allow for violations of the entropy bound during the evaporation process. In our picture, information about the star's interior is not carried by the Hawking particle (nor do we rely on the existence of a physical stretched horizon as a membrane), and it can only be distinguished after the entire evaporation process has occurred. In contrast, black hole complementarity suggests that Hawking radiation should carry information. The reason why black hole complementarity says Hawking radiation carries information is due to the entropy bound relation \cite{Page:1993wv}. However, if that is not the case, there is no reason to force Hawking radiation to carry information. This resolves the potential inconsistency of the original version of black hole complementarity.

One might think that the correspondence between the Boltzmann entropy and the Bekenstein-Hawking entropy is appropriate for the case of thermal equilibrium (for eternal black holes). On the other hand, if the entropy-violating effects accumulate, then the system becomes a non-equilibrium state and, in this limit, there may appear a discrepancy between the Bekenstein-Hawking entropy and the Boltzmann entropy. However, after the total evaporation, if all particles and antiparticles appear in a causally connected region, the number of states might be reduced \cite{Hwang:2017yxp}, which is consistent with the entropy bound relation after a sufficient thermalization process.


\end{document}